\documentclass[a4paper,11pt]{article}
\usepackage{pos}
\usepackage{mathrsfs}
\usepackage{slashed}

\newcommand{\U}{\mathcal{U}}
\newcommand{\Z}{\mathcal{Z}}
\newcommand{\pine}{\mathcal{P}}
\newcommand{\PP}{\mathbb{P}}
\newcommand{\Tr}{\mathrm{Tr}}
\newcommand{\muI}{\mu_I}
\newcommand{\Chat}{\widehat{\mathcal{C}}}
\newcommand{\cV}{\mathcal{V}}
\newcommand{\cT}{\mathcal{T}}
\newcommand{\cJ}{\hat{\mathcal{J}}}
\newcommand{\Mud}{\mathcal{M}_{ud}}
\newcommand{\Msv}{\mathscr{M}}

\title{Leading-order dynamical QED corrections to the thermodynamics of isospin-asymmetric QCD}
\ShortTitle{Dynamical QED corrections to isospin-asymmetric QCD}

\renewcommand{\speakerText}{Speaker (intended); the talk was presented by G.~Mark\'o.}

\author[a]{Gergely Endr\H{o}di}
\author[a]{\'Ad\'am Fischer}
\author*[a]{Wei-Ping Huang}
\author[a]{Gergely Mark\'o}

\affiliation[a]{Institute of Physics and Astronomy, ELTE E\"otv\"os Lor\'and University,\\
  P\'azm\'any P\'eter s\'et\'any 1/A, H-1117 Budapest, Hungary}

\emailAdd{huang.weiping@ttk.elte.hu}

\abstract{
We present a first lattice calculation of the leading-order effect of dynamical photons on the thermodynamics of QCD at nonzero isospin chemical potential $\muI$, across the onset of charged-pion condensation. 
In QCD this onset is a second-order transition, which dynamical photons are expected, on general grounds, to drive first order. However, a direct simulation is not possible since unequal quark charges render the QCD+QED action complex at nonzero $\muI$. We therefore expand the free energy density $f$ in the electromagnetic coupling $e$ around $e=0$, where the sign problem is absent. 
The leading-order expansion coefficient, $\partial^2 f/\partial e^2|_{e=0}$, is the convolution of the photon propagator with the electromagnetic current--current correlator, which we measure with a newly proposed planar source on existing (2+1)-flavor pure-QCD ensembles at nonzero $\muI$ and pionic source $\lambda$. 
At physical quark masses and a single lattice spacing, after extrapolating to vanishing $\lambda$, we find that $\partial^2 f/\partial e^2|_{e=0}$ is, within errors, independent of $\muI$ below the onset and rises sharply across it.}

\notes{\note[]{The authors are grateful to Eduardo Garnacho-Velasco for discussions. This research received funding from the Hungarian National Research, Development and Innovation Office - NKFIH (Research Grant Hungary 150241) and the European Research Council (Consolidator Grant 101125637 CoStaMM).}}

\FullConference{The 43rd International Symposium on Lattice Field Theory (Lattice 2026)\\
July 26 to August 1, 2026\\
University of Maryland, College Park, USA\\}

\begin{document}

\maketitle

\section{Introduction}\label{sec:intro}

Strong-interaction matter at nonzero isospin chemical potential $\muI$ is relevant for the description of systems with an isospin asymmetry, e.g., heavy-ion collisions~\cite{Li:2008gp}, compact stars~\cite{Steiner:2004fi} and the early Universe~\cite{Wygas:2018otj}.
Unlike nonzero baryon density, this regime is accessible to lattice simulations, since for $\mu_u=-\mu_d=\muI$ the $\gamma_5$-hermiticity of the Dirac operator pairs the two light flavors and the fermion determinant is thus real and non-negative~\cite{Kogut:2002zg,Brandt:2017oyy}.
The phase diagram in the $T$--$\muI$ plane has therefore been studied in detail. At low temperature the system undergoes a second-order transition at $\muI\sim m_\pi/2$ into a Bose--Einstein condensate (BEC) of charged pions~\cite{Son:2000xc,Brandt:2017oyy}, while at high temperature the hadronic phase turns into the quark--gluon plasma through a crossover.
Since spontaneous symmetry breaking does not occur in a finite volume, a pionic source $\lambda$ is additionally needed to regulate the infrared behavior of the condensed phase and expose the order parameter. To restore the physical situation, this auxiliary parameter has to be extrapolated to zero at the end.
In this way the BEC onset has been located with continuum-extrapolated staggered quarks at physical masses, around which the pion condensate was shown to follow O(2) scaling~\cite{Brandt:2017oyy}.

The picture changes once the dynamical photon field is involved. The condensing pions are charged and couple to the photon through the covariant derivative, so that in the condensed phase the photon acquires a mass $m_A\propto e|\pi^\pm|$, which gives rise to the Meissner effect as in a superconductor. 
Near a continuous transition at nonzero temperature, the long-distance behavior is described by a three-dimensional effective theory of the bosonic zero Matsubara modes, in which integrating out the photon of mass $m_A$ contributes $\propto-m_A^3$ to the effective potential. 
Such a non-analytic cubic contribution $\propto-e^3|\pi^\pm|^3$ cannot appear in a Landau expansion in powers of $|\pi^\pm|^2$. 

The photon-induced cubic term underlies the Halperin--Lubensky--Ma mechanism~\cite{Halperin:1973jh}, discussed also for color superconductivity~\cite{Giannakis:2004xt}. 
Without it, the minimum of the potential moves away from zero continuously, while with it a barrier and a rival minimum appear and the condensate jumps. 
Including dynamical photons can therefore drive the BEC onset first order, which calls for a first-principles determination of the order of the pion condensation transition in QCD+QED.

Direct simulations of QCD+QED at $\muI\neq0$ suffer from a sign problem, as explained in Sec.~\ref{sec:pineapple}. We instead expand the free energy density $f$ around $e=0$. 
Its leading expansion coefficient, $\partial^2f/\partial e^2|_{e=0}$, is a convolution of the photon propagator with the electromagnetic current--current correlator of pure QCD and can therefore be computed on sign-problem-free QCD ensembles at nonzero $\muI$ and $\lambda$.
Such an expansion yields the non-analytic cubic term only after a resummation of the photon self-energy, but its leading order already shows where along the $\muI$ axis and how strongly QED acts on the thermodynamics.
Sec.~\ref{sec:lattice} discusses the simulations with a new two-dimensional planar source and an issue with it which we call geometric misalignment.
Sec.~\ref{sec:results} presents a first lattice determination of the coefficient as a function of $\muI$ across the onset, and Sec.~\ref{sec:conclusions} concludes.

\section{Expansion in \texorpdfstring{$e$}{e} around QCD at \texorpdfstring{$\muI\neq0$}{muI<>0}: the pineapple}\label{sec:pineapple}

We consider (2+1)-flavor QCD in the rooted staggered discretization, with degenerate up and down quark masses $m_u=m_d\equiv m_l$ and a strange quark mass $m_s$. 
With a nonzero isospin chemical potential $\mu_u=-\mu_d=\muI$ and a pionic source $\lambda$ for the light flavors, this theory becomes isospin-asymmetric but remains free of the sign problem. 
The staggered Dirac operators of the light flavors, $M_{u,d}=\slashed{D}(\mu_{u,d})+m_l$, are related through the $\eta_5$-hermiticity $\eta_5 M_u^\dagger\eta_5=M_d$, keeping the determinant of the light matrix $\Mud=\mathrm{diag}(M_u,M_d)+i\lambda\eta_5\tau_2$ real and positive~\cite{Kogut:2002zg,Brandt:2017oyy}. 

However, once the electromagnetic coupling $e$ is turned on, dynamical photons couple to each flavor $f$ through its charge $q_{f}$. 
Because $q_u\neq q_d$, the $\eta_5$-hermiticity between $M_u$ and $M_d$ is broken, leading to a complex QCD+QED action at $\muI\neq0$. 
Considering that the QED coupling is weak, we bypass the sign problem by expanding the free energy density in $e$ around pure QCD (i.e., $e=0$), as done, for example, for the isospin-breaking corrections computed in Ref.~\cite{deDivitiis:2013xla}.

To set up the expansion, we start from the weight of an SU(3) gauge configuration $\U$ at $e\neq0$ as $w(\U;\muI;eA)=\exp\{-S_g[\U]\}\det\Mud^{1/4}\det M_s^{1/4}$, with $S_g$ the gauge action and $M_s=\slashed{D}(\mu_s=0)+m_s$ the strange Dirac operator. 
The photon field $A_{\mu}$ enters the Dirac operator $M_{f}$ through the gauge link $U_\nu(x)\to \exp\{ieq_fA_\nu(x)\}U_\nu(x)$. 
We then expand the weight to second order in the photon field on a fixed configuration,
\begin{equation}
  w(\U;\muI;eA)=w_0\Big[1+ie\sum_{x,\mu}A_\mu(x)\,\cJ_\mu(x)-\frac{e^2}{2}\sum_{x,y,\mu,\nu}A_\mu(x)A_\nu(y)\,\Chat_{\mu\nu}(x,y)+\mathcal O(e^3)\Big]\,,
  \label{eq:wexp}
\end{equation}
where $w_0\equiv w(\U;\muI;eA=0)$ is the pure-QCD weight. 
The coefficients $\cJ_\mu(x)$ and $\Chat_{\mu\nu}(x,y)$ are the one-point current and current--current operators, given explicitly in Sec.~\ref{sec:ensembles}.

Since the photon action is Gaussian, the photon field average $\langle\cdot\rangle_\gamma\equiv \int\mathcal{D}A e^{-S_\gamma}(\cdot)\big/{\int\mathcal{D}A e^{-S_\gamma}}$ eliminates all odd moments of $A_{\mu}$ and gives the free photon propagator $\langle A_\mu(x)A_\nu(y)\rangle_\gamma=\Delta^{(0)}_{\mu\nu}(x,y)$. 
Performing the photon integral in the partition function $\Z=\int\mathcal{D}\U\!\int\mathcal{D}Ae^{-S_\gamma}w(\U;\muI;eA)$ gives $\Z=\Z|_{e=0}\big[1-\tfrac{e^2}{2}\sum_{x,y,\mu,\nu}\Delta^{(0)}_{\mu\nu}(x,y)\langle\Chat_{\mu\nu}(x,y)\rangle_\U+\mathcal{O}(e^4)\big]$, with $\langle\cdot\rangle_\U\equiv\int\mathcal{D}\U w_0(\cdot)\big/\int\mathcal{D}\U w_0$ the average over the pure-QCD ensemble.
Hence for $f=-V_4^{-1}\ln \Z$ on the $V_4=N_s^3N_\tau$ lattice,
\begin{equation}
  f(\muI,e)=f(\muI,0)+\frac{e^2}{2}\pine+\mathcal O(e^4)\,,
  \text{with}\;
  \pine\equiv\frac{\partial^2 f}{\partial e^2}\bigg|_{e=0}
  =\frac{1}{V_4}\sum_{x,y,\mu,\nu}\Delta^{(0)}_{\mu\nu}(x,y)\,\big\langle\Chat_{\mu\nu}(x,y)\big\rangle_\U\,.
  \label{eq:pineapple-def}
\end{equation}
The $\mathcal O(e^2)$ coefficient $\pine$, which we call the `pineapple' after the shape of its diagram, is a photon propagator tying together two current insertions on the quark loop. 
It is evaluated on the sign-problem-free pure-QCD ensembles, and the photon part of the diagram is known analytically. 
The derivative in Eq.~(\ref{eq:pineapple-def}) is taken at fixed bare gauge coupling $\beta$ and quark masses, where we currently ignore their dependence on $e$ along a line of constant physics.

In practice, we work in Feynman gauge with the lattice photon propagator at the same finite volume and lattice spacing as the QCD ensemble itself, i.e., $\Delta^{(0)}_{\mu\nu}(x,y)=\delta_{\mu\nu}D(x-y)$ with $D(x)=V_4^{-1}\sum_{k\in{\rm BZ}\cap\Msv_P}\exp\{ik\cdot x\}/\hat{k}^2$ and $\hat{k}^2=\sum_\mu 4\sin^2(k_\mu/2)$, where BZ denotes the Brillouin zone. 
The massless propagator is singular at the zero mode, so an infrared prescription $\Msv_P$ is mandatory. ${\rm QED}_L$~\cite{Hayakawa:2008an} drops all modes with $\vec k=0$, giving $\Msv_{\rm L}=\{k:\vec k\neq0\}$, whereas ${\rm QED}_{TL}$~\cite{Duncan:1996xy} drops only the global one, with $\Msv_{\rm TL}=\{k:k\neq0\}$. 
By translation invariance the pineapple in Eq.~(\ref{eq:pineapple-def}) reduces to a single sum in position space over the traced correlator $\mathcal{C}(x)\equiv \sum_\mu \mathcal{C}_{\mu\mu}(x)$ with $\mathcal{C}_{\mu\nu}(x)\equiv\langle\Chat_{\mu\nu}(x,0)\rangle_\U$, or, equivalently, over the traced polarization in the BZ,
\begin{equation}
  \pine=\sum_x D(x)\,\mathcal{C}(x)
  =\frac{1}{V_4}\sum_{k\in{\rm BZ}\cap\Msv_P}\frac{\Pi(k)}{\hat k^2}\,,
  \quad\text{with}\quad
  \Pi(k)=\sum_x e^{-ik\cdot x}\,\mathcal{C}(x)\,.
  \label{eq:pineapple}
\end{equation}

The pineapple $\pine$ is dominated by the short-distance contributions, which depend on the regularization scheme but not on the state of the medium (i.e., $\muI$-independent).
The $\muI$ dependence, by contrast, is expected to be governed by the infrared physics, and it is what we concentrate on.
Furthermore, the pineapple determines the leading-order QED effects on the QCD thermodynamics. For example, since the pressure is $p=-f$, dynamical photons shift it at order $e^2$ by $-\frac{e^2}{2}\pine$ and the isospin density $n_I=\partial p/\partial\muI$ by $-\frac{e^2}{2}\partial\pine/\partial\muI$, which will be covered in our future work.

\section{Lattice simulation}\label{sec:lattice}

\subsection{Setup and observable}\label{sec:ensembles}

We use the isospin-asymmetric ensembles of Ref.~\cite{Brandt:2017oyy}, generated with $N_f=2+1$ staggered quarks with stout smearing at physical pion mass and a tree-level Symanzik-improved gauge action~\cite{Borsanyi:2010cj} on $N_s^3\times N_\tau=24^3\times6$ lattices of a single lattice spacing $a$ with $\beta=3.45$.
These are pure-QCD configurations without any photon field, as the strategy of Sec.~\ref{sec:pineapple} requires. 
We scan six chemical potentials $a\muI$ between 0.025 and 0.15 covering the BEC onset region around $a\muI^{\rm crit}\approx0.1$, and four pionic sources $a\lambda$ between 0.0006 and 0.0075, with four values of $a\muI$ at each source.

The observable is the full (2+1)-flavor current--current operator $\langle\Chat_{\mu\nu}(x,y)\rangle_\U$ defined in Eq.~(\ref{eq:wexp}).
With the current and tadpole vertices $\cV^f_\mu(x)\equiv(ieq_f)^{-1}{\partial M_f}\big/{\partial A_\mu(x)}\big|_{A=0}$ and $\cT^f_{\mu\nu}(x)\equiv(ieq_f)^{-2}{\partial^2M_f}\big/{\partial A_\mu(x)\partial A_\nu(x)}\big|_{A=0}$ detailed in Sec.~\ref{sec:planar}, one finds
\begin{equation}
  \Chat_{\mu\nu}(x,y)
  =\frac{\delta_{x,y}}{4}\Tr\big[\Mud^{-1}\mathbf Q^2\cT_{\mu\nu}(x)\big]
  -\frac{1}{4}\Tr\big[\Mud^{-1}\mathbf Q\cV_\mu(x)\,\Mud^{-1}\mathbf Q\cV_\nu(y)\big]
  +(ud\to s)
  +\cJ_\mu(x)\cJ_\nu(y)\,,
  \label{eq:Chat}
\end{equation}
with $\cJ_\mu(x)=\frac14\Tr[\Mud^{-1}\mathbf Q\cV_\mu(x)]+\frac{q_s}{4}\Tr[M_s^{-1}\cV^s_\mu(x)]$, $\mathbf Q=\mathrm{diag}(q_u,q_d)$, $\cV_\mu=\mathrm{diag}(\cV^u_\mu,\cV^d_\mu)$ and $\cT_{\mu\nu}=\mathrm{diag}(\cT^u_{\mu\nu},\cT^d_{\mu\nu})$. 
The traces run over flavor, site and color, and $(ud\to s)$ denotes the strange tadpole and connected terms, obtained by substituting $\Mud^{-1}\mathbf Q\to q_sM_s^{-1}$ and using the strange vertices. We measure the tadpole, quark-line connected and disconnected terms of Eq.~(\ref{eq:Chat}), except for the cross-flavor term proportional to $\lambda^2$ in the light connected piece.
Using $\eta_5M_u^\dagger\eta_5=M_d$, all light traces reduce to traces of $G_u\equiv S_uM_u^\dagger$ with $S_u\equiv(M_u^\dagger M_u+\lambda^2)^{-1}$; the relevant light solves use a conjugate-gradient algorithm accelerated by projecting out the lowest eigenmodes of $M_u^\dagger M_u$. 

Note that $f$ in Eq.~(\ref{eq:pineapple-def}) is given by the logarithm of $\Z$ after the photon integration.
Taking the logarithm at fixed photon field and averaging afterwards would instead replace $\langle\Chat_{\mu\nu}(x,y)\rangle_\U$ by $\langle\Chat_{\mu\nu}(x,y)\rangle_\U-\langle\cJ_\mu(x)\rangle_\U\langle\cJ_\nu(y)\rangle_\U$, since the photon average does not commute with the logarithm, i.e., $\langle\ln\int\mathcal{D}\U\,w\rangle_\gamma\neq\ln\langle\int\mathcal{D}\U\,w\rangle_\gamma$.
At $\muI\neq0$ only the charge density $\langle\cJ_4\rangle_\U$ is nonzero because $\muI$ couples to the temporal component and breaks charge conjugation, and by translation invariance the two orderings differ by $\langle\cJ_4\rangle_\U^2\sum_xD(x)$ in Eq.~(\ref{eq:pineapple}).
This difference vanishes for ${\rm QED}_L$ and ${\rm QED}_{TL}$, whose propagators have no zero mode, so that the product of one-point functions can be subtracted without changing $\pine$.
We subtract this product in what follows, which matters only for the interpolation of $\Pi(k)$ in Sec.~\ref{sec:bz}, because it enters only $\Pi(0)$ and would distort the profile.

\subsection{The planar source, geometric misalignment and its treatment}\label{sec:planar}

Let $(\PP_\zeta)_{nm}=\delta_{nm}\,\delta_{n\in{\rm Plane}_\zeta}$ be the projector onto the $x$--$y$ plane with normal coordinates $\zeta=(z,\tau)$, and $\gamma_\mu\equiv\sum_x\cV_\mu(x)$ the lattice-summed current vertex defined in Sec.~\ref{sec:ensembles}.
Summing the currents over the source $(\zeta')$ and sink $(\zeta)$ planes defines the two-dimensional correlator
\begin{equation}
  \mathcal C^{\rm 2D}_{\mu\nu}(\Delta z,\Delta\tau)
  =\frac{1}{N_\perp}\sum_{x\in{\rm Plane}_\zeta} \sum_{y\in{\rm Plane}_{\zeta'}} \langle\Chat_{\mu\nu}(x,y)\rangle_\U
  =\sum_{\vec x_\perp}\mathcal C_{\mu\nu}(\vec x_\perp,\Delta z,\Delta\tau)\,,
  \label{eq:C2D}
\end{equation}
with $(\Delta z,\Delta\tau)=\zeta-\zeta'$ and $N_\perp\equiv N_s^2$ the number of sites per plane.
For an in-plane current both ends of every bond lie in the plane, so the plane sum is simply $\sum_{x\in{\rm Plane}_\zeta}\cV_\mu(x)=\gamma_\mu\PP_\zeta$.
In a connected light trace such as $\Tr[G_u\gamma_\mu\PP_\zeta\,G_u\gamma_\nu\PP_{\zeta'}]$, the source projector is realized by a planar source, a Gaussian noise restricted to the source plane, $\xi_{\zeta'}=\PP_{\zeta'}\xi$ with $\langle\xi_{\zeta'}\xi^\dagger_{\zeta'}\rangle_\xi=\PP_{\zeta'}$, in a one-end trick~\cite{McNeile:2002fh}.
The sink projector only restricts the site sum, so one set of inversions per noise serves all sink planes.
The double plane sum projects to zero in-plane momentum, so the two-dimensional Fourier transform of $\mathcal C_{\mu\nu}^{\rm 2D}$ gives the polarization exactly on the axis slice $k_1=k_2=0$, from which the Brillouin-zone sum of Eq.~(\ref{eq:pineapple}) is reconstructed in Sec.~\ref{sec:bz}.

For a crossing current with $\hat\mu$ along one of the normals, every bond connects two neighboring planes, and $\gamma_\mu\PP_\zeta$ is not the plane sum.
The vertex splits into a forward and a backward hop along the bond $(x,x+\hat\mu)$, $[\cV^{(+)}_\mu(x)]_{nm}\propto U_\mu(x)\,\delta_{n,x}\delta_{m,x+\hat\mu}$ and $[\cV^{(-)}_\mu(x)]_{nm}\propto U^\dagger_\mu(x)\,\delta_{n,x+\hat\mu}\delta_{m,x}$, and only the two together form the conserved current on that bond.
With $\gamma^{(\pm)}_\mu\equiv\sum_x\cV^{(\pm)}_\mu(x)$, an in-plane hop stays in the plane, $[\gamma^{(\pm)}_\mu,\PP_\zeta]=0$, whereas a crossing hop shifts it,
\begin{equation}
  \gamma^{(+)}_\mu\PP_{\zeta+\hat\mu}=\PP_\zeta\,\gamma^{(+)}_\mu\,,\qquad
  \gamma^{(-)}_\mu\PP_{\zeta}=\PP_{\zeta+\hat\mu}\,\gamma^{(-)}_\mu\,.
  \label{eq:geo-comm}
\end{equation}
The plane sum of the bond current therefore projects the two hops onto different planes,
\begin{equation}
  \sum_{x\in{\rm Plane}_\zeta}\cV_\mu(x)=\PP_\zeta\,\gamma^{(+)}_\mu+\PP_{\zeta+\hat\mu}\,\gamma^{(-)}_\mu
  =\gamma^{(+)}_\mu\PP_{\zeta+\hat\mu}+\gamma^{(-)}_\mu\PP_\zeta\,,
  \label{eq:plane-sum}
\end{equation}
whereas the planar source implements $\gamma_\mu\PP_\zeta=\PP_{\zeta-\hat\mu}\gamma^{(+)}_\mu+\PP_{\zeta+\hat\mu}\gamma^{(-)}_\mu$, which takes the forward hop from the bonds originating on the plane behind.
The two halves of this current lie one lattice unit apart and land in bins of the normal separation that differ by one unit, so that the measured correlator no longer satisfies Eq.~(\ref{eq:C2D}). 
The same misalignment affects the crossing currents of any source restricted to a sublattice.

We resolve this geometric misalignment with a point-split one-end construction that realizes Eq.~(\ref{eq:plane-sum}), keeping the forward and backward source halves apart and binning each by the position of its bond, at the cost of a third inversion per noise for a crossing current.
The plane-summed one-point functions of the disconnected piece, with the same bond assignment, are obtained on all parallel planes at once from noise on the whole lattice, with the plane projection on the sink side. Their products are formed from independent noises using all cross pairs, so that no self-contraction appears, and the tadpole vertex reuses the same solution fields.

We checked that the correlators measured with the planar source agree within errors, in all connected, disconnected and tadpole channels, with the point-source correlators projected according to Eq.~(\ref{eq:C2D}) and with the one-dimensional projection from a wall source~\cite{Bali:2015msa} on the same ensemble.

\section{Results}\label{sec:results}

\subsection{From the planar correlator to the Brillouin-zone sum}\label{sec:bz}

The pineapple, Eq.~(\ref{eq:pineapple}), needs the polarization at every momentum of the BZ, whereas the planar correlator fixes it exactly only on the axis slice. A two-dimensional Fourier transform of the measured $\mathcal C^{\rm 2D}(\Delta z,\Delta\tau) \equiv\sum_{\mu} \mathcal C^{\rm 2D}_{\mu\mu}(\Delta z,\Delta\tau)$ gives the knots of the interpolation
\begin{equation}
  Y_{n\ell}\equiv\sum_{\Delta z,\Delta\tau}e^{-i(q_n\Delta z+\omega_\ell\Delta\tau)}\,\mathcal C^{\rm 2D}(\Delta z,\Delta\tau)=\Pi(0,0,q_n,\omega_\ell)\,,\;
  \text{with}\quad
  q_n=\frac{2\pi n}{N_s}\,,\;
  \omega_\ell=\frac{2\pi\ell}{N_\tau}\,,
  \label{eq:knots}
\end{equation}
where $n=0,\dots,N_s/2$ and $\ell=0,\dots,N_\tau/2$, the three spatial axes being equivalent.

The remaining momenta are reconstructed from the spatial O(3) symmetry of the medium, i.e., $\Pi(k)=\Pi(\omega,|\vec k|)$, which turns the on-axis data into a radial profile $F_\ell(q^2)$ at each Matsubara frequency. 
On the lattice, however, this is an approximation, since O(3) is broken to the cubic group. The on-axis data cannot distinguish cubic-invariant deviations such as $\hat k_1^2\hat k_2^2+\hat k_2^2\hat k_3^2+\hat k_3^2\hat k_1^2$, whose relative effect is of order $a^2q^2$ at small momentum but of order one near the BZ boundary.
In practice, the profile is assigned per shell $s=a^2+b^2+c^2$ of equal $q=(2\pi/N_s)\sqrt s$, whereas the photon weight is constant only on a cubic orbit $\mathcal O_{abc}$ of integer triples with $a\ge b\ge c$, and is accumulated per orbit, 
\begin{equation}
  K_{abc,\ell}=\frac{g_{abc}}{V_4\,(\hat q^2_{abc}+\hat\omega^2_\ell)}\,,\quad
  \pine\simeq\sum_{(abc,\ell)\in{\rm BZ}\cap\Msv_P}K_{abc,\ell}\;{\rm Re}\,F_\ell\big(q^2_{abc}\big)\,,
  \label{eq:assembly}
\end{equation}
with $g_{abc}\equiv|\mathcal O_{abc}|$ the number of distinct lattice momenta in the orbit. 

\begin{figure}[ht]
  \centering
  \includegraphics[width=0.45\textwidth]{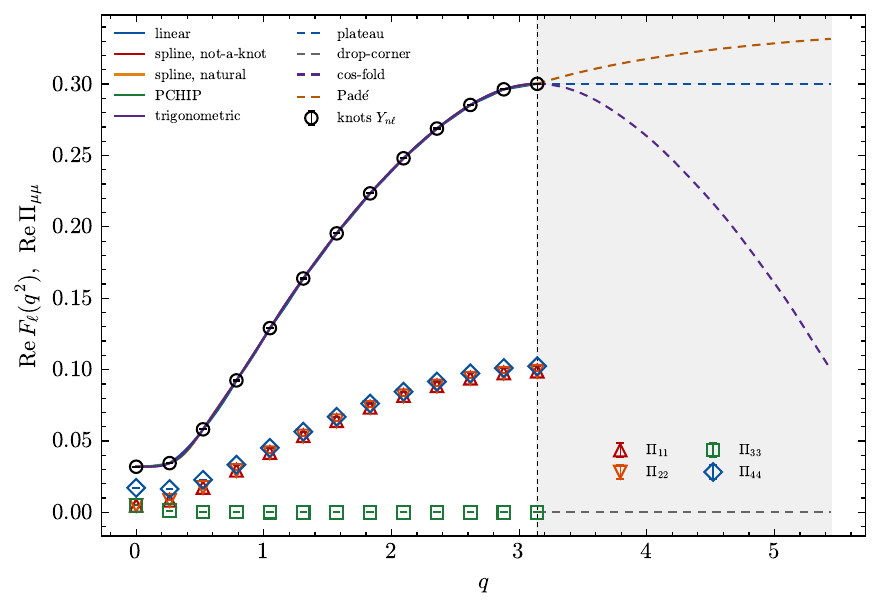}\hfill
  \includegraphics[width=0.45\textwidth]{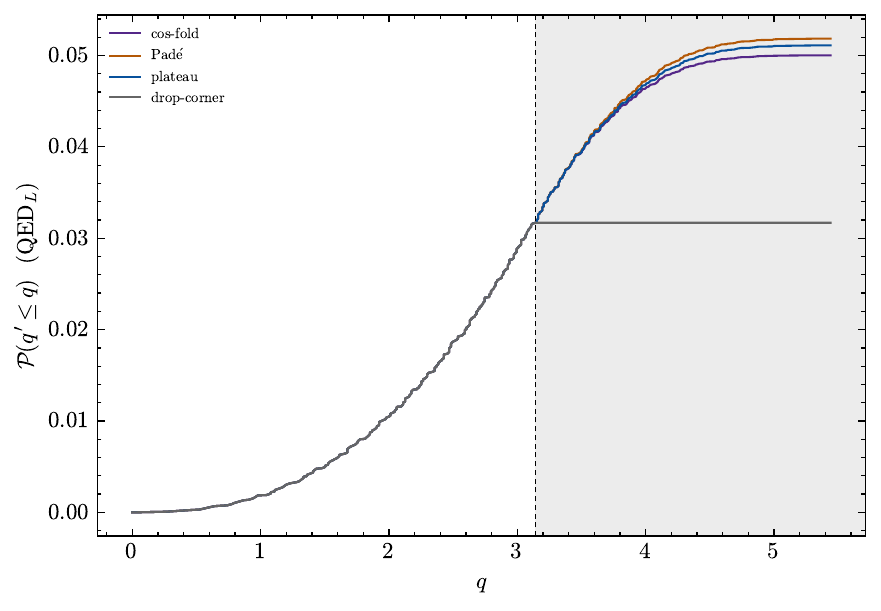}
  \caption{Polarization reconstruction at $a\muI=0.1$, $a\lambda=0.0025$ with ${\rm QED}_L$. Left: the $\ell=0$ radial profile ${\rm Re}\,F_0$ as a function of $q$, with the measured knots (circles), the interior interpolations for $q\le\pi$ and the four corner continuations in the shaded region $q>\pi$. The colored points are the diagonal components ${\rm Re}\,\Pi_{\mu\mu}$. Right: the Brillouin-zone sum accumulated in order of increasing $q$; its right edge is the pineapple.}
  \label{fig:recon}
\end{figure}

Two regions must be distinguished. In the interior, $q\le\pi$, the profile is fixed by the knots and off-knot radii are interpolated in $q^2$. As shown in the left panel of Fig.~\ref{fig:recon}, different interpolants through all knots are indistinguishable, so that the interpolation adds no visible uncertainty there.

The corner of the BZ, $\pi<q\le\sqrt3\,\pi$, contains no data, and we compare four model completions, namely a plateau at the edge value $Y_{N_s/2,\ell}$ (plateau), the periodic continuation of the trigonometric interpolant (cos-fold), a Pad\'e form $c_0+c_1/(q^2+c_2)$ fitted to the eight outermost knots and anchored continuously at $q=\pi$ (Pad\'e), and the truncation $F_\ell=0$ (drop-corner).

Fig.~\ref{fig:recon} shows the reconstruction on the ensemble at $a\muI=0.1$, $a\lambda=0.0025$, taking $\ell=0$ as an example. The left plot presents the interpolations and continuations in the two regions of the BZ, while the right one gives the BZ sum accumulated in order of increasing $q$, whose right edge is the pineapple.
In the right plot all schemes coincide in the interior, and the interpolation spread of the result stays below 0.1\%.
At $q=\pi$ the curves split. The drop-corner line stays flat by construction, while the plateau, Pad\'e and cos-fold lines keep rising and end within a few percent of each other.
The BZ corner thus supplies nearly 40\% of the pineapple, entirely from the completion model. The results below are therefore given for the plateau and drop-corner completions for comparison. 

The $q\to0$ limit of $\mathrm{Re}\,F_0$ can be linked to the dynamically generated screening masses of the photon through the static diagonal polarization $\Pi_{\mu\mu}(0,q)$, among which the longitudinal $\Pi_{33}$ (for momenta along the third axis) vanishes for $q\neq0$ by the Ward identity, which, like the relations below, holds only for $\lambda\to0$.
In the normal phase only the electric component remains, $e^2\,{\rm Re}\,F_0(q\to0)=e^2\,{\rm Re}\,\Pi_{44}(0,q\to0)=m_D^2$ with $m_D$ the Debye mass, while in the condensed phase the two transverse components add $2m_A^2$ with $m_A$ the Meissner mass.
Since $m_D>0$ at any $T>0$ (and even at $T=0$ in the condensed phase), the limit is nonzero at all $\muI$ even for vanishing source, but it combines two qualitatively different mechanisms, which the separately measured $\Pi_{44}$ and $\Pi_{11}+\Pi_{22}$ distinguish, as needed for the non-analytic term $\propto m_A^3$ in Sec.~\ref{sec:intro}.

\subsection{Extrapolation to vanishing pionic source}\label{sec:lambda0}

\begin{figure}[h]
  \centering
  \includegraphics[width=0.45\textwidth]{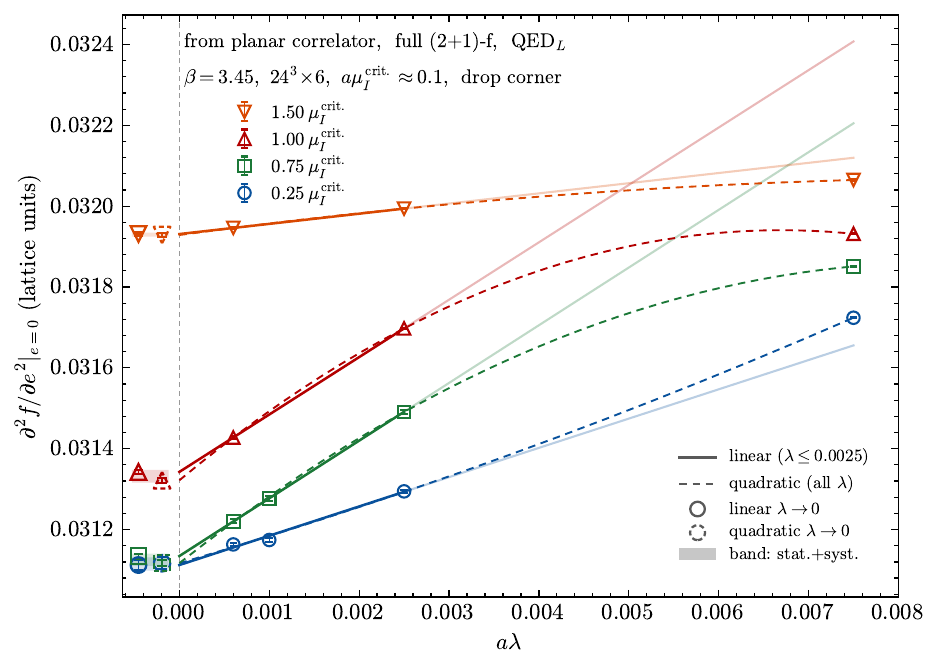}
  \caption{Pineapple as a function of the pionic source at four isospin chemical potentials, quoted in units of $a\muI^{\rm crit}\approx0.1$, for the drop-corner completion. Solid lines are the linear fits to $a\lambda\le0.0025$, dashed lines the quadratic fits through all points; the markers left of $a\lambda=0$ are the two extrapolated values and the bands their envelopes. The plateau completion gives the same picture with a larger absolute value.}
  \label{fig:lambda0}
\end{figure}

The pionic source is not physical and, since it carries the charge of the pion, also breaks the electromagnetic U(1) explicitly, so that the pineapple $\pine$ is gauge dependent at finite $\lambda$. 
The results are therefore extrapolated at each $\muI$ to $\lambda=0$,  where the electromagnetic current is exactly conserved and $\pine$ is gauge independent. 
As shown in Fig.~\ref{fig:lambda0}, the largest source $a\lambda=0.0075$ deviates visibly from the linear trend of the smaller ones. 
We therefore take as the central value a linear fit $A+B\lambda$ restricted to $a\lambda\le0.0025$, and a quadratic $A+B\lambda+C\lambda^2$ through all points for comparison. The band at $\lambda=0$ is the envelope of the two fits. 
Both the linear and quadratic extrapolations are empirical and based on limited data, and at $a\muI=0.1$ and $0.15$ only three source values exist, so that neither fit has a residual degree of freedom. The band therefore reflects the spread between two empirical forms, not a complete systematic error. 
A more reliable extrapolation, based on the improvement through the singular-value representation of the Dirac operator~\cite{Brandt:2017oyy}, is in progress. 

\subsection{Dependence on the isospin chemical potential}\label{sec:muI}

\begin{figure}[ht]
  \centering
  \includegraphics[width=0.45\textwidth]{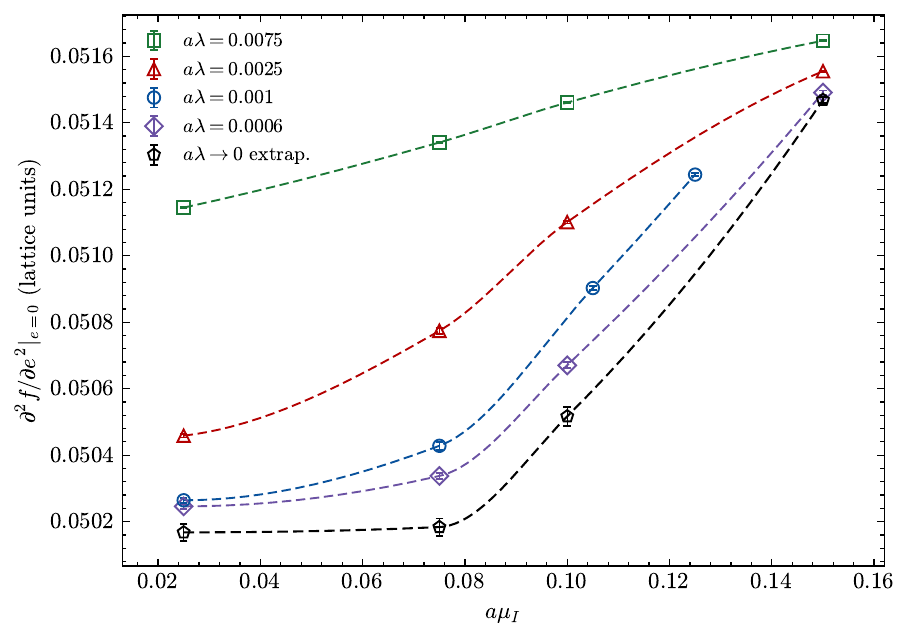}\hfill
  \includegraphics[width=0.45\textwidth]{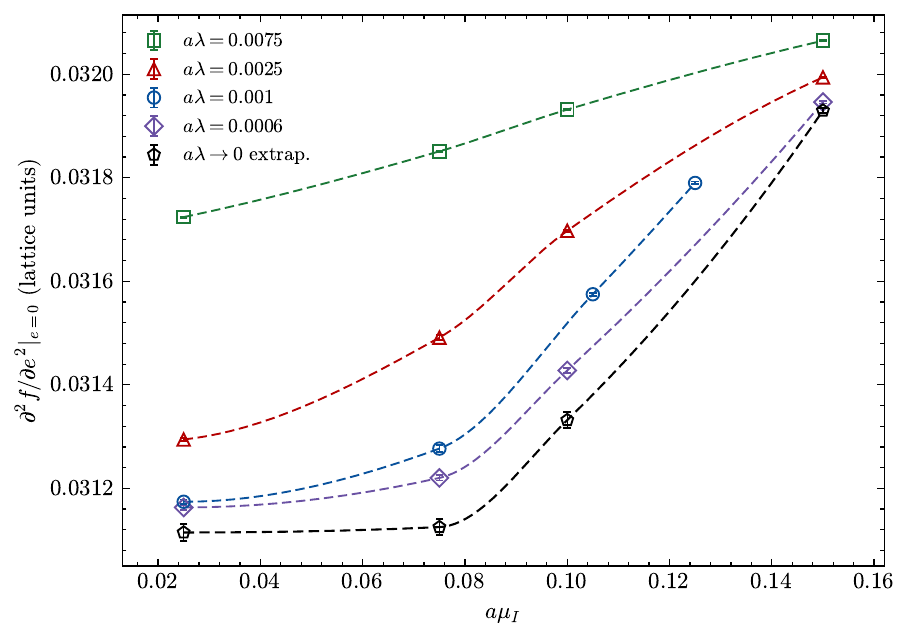}
  \caption{Pineapple as a function of $a\muI$ for the four pionic sources (colors) and after the extrapolation $\lambda\to0$ (black), with the plateau (left) and the drop-corner (right) completion of the Brillouin-zone corner. Lines are drawn to guide the eye.}
  \label{fig:muI}
\end{figure}

Fig.~\ref{fig:muI} collects the pineapple on the whole grid for the plateau and drop-corner completions, together with the extrapolated values. 
The two panels differ by the corner contribution of Sec.~\ref{sec:bz}, but show the same qualitative picture, i.e., a plateau below the onset and a sharp rise across it at the same location, while the size of the rise differs by more than a third.

At finite $\lambda$ all curves rise monotonically over the whole range, a larger source giving a larger value and a flatter curve, and the curves for different $\lambda$ approach each other at the largest $\muI$. 
Only after the extrapolation does a plateau appear, the extrapolated value being flat within errors below $a\muI^{\rm crit}\approx0.1$. 
This is the behavior expected from the Silver Blaze property~\cite{Cohen:2003kd}, which states that below the threshold the creation of a charged pion costs energy, so that the ground state and the thermodynamics, including its QED correction, do not depend on $\muI$. 
Strictly this holds at zero temperature, and at our $T\approx113$~MeV the plateau here is Silver-Blaze-like. 
At the onset the extrapolated curve turns sharply upward and keeps rising.

\section{Conclusions and outlook}\label{sec:conclusions}

We have computed the leading dynamical-QED correction to the free energy density of isospin-asymmetric QCD, the pineapple $\partial^2f/\partial e^2|_{e=0}$, at the physical point across the pion-condensation onset. 
The (2+1)-flavor current--current correlator, including connected, disconnected and tadpole contributions, was measured with a novel two-dimensional planar source, whose geometric misalignment is removed by a point-split treatment.
These planar data fix the polarization exactly on an axis slice of the Brillouin zone, so that the full momentum sum had to be reconstructed, by interpolation in the interior and by model completions in the unmeasured corner. 
After an empirical extrapolation of the pionic source to zero, the pineapple is independent of $\muI$ within errors below the onset and rises sharply across it, qualitatively in the same way for both corner completions, while the size of the rise depends on the completion.
This indicates that dynamical QED feels the condensation of charged pions already at leading order, which is the behavior required if photons are to affect the order of the transition.

The present results are a first estimate of the pineapple at a single lattice spacing. Its renormalization and the continuum limit are left for future work. 
In addition, we find that point-source estimators give a signal comparable to that of the planar source. With the full four-dimensional correlator they provide, the corner systematics can be eliminated in future work, allowing a complete Brillouin-zone sum without any reconstruction.

\bibliographystyle{JHEP.bst}
\bibliography{ref.bib}

\end{document}